\documentclass[11pt,sort&compress]{elsarticle}

\journal{J.\ Comp.\ Phys.}
\usepackage{bm}
\newcommand{\ve}[1]{\bm{#1}}

\newcommand{\bu}{\ve{u}}

\newcommand{\bw}{\ve{w}}
\newcommand{\by}{\ve{y}}
\newcommand{\bx}{\ve{x}}

\newcommand{\bJ}{\ve{J}}

\newcommand{\bV}{\ve{V}}
\newcommand{\bW}{\ve{W}}
\newcommand{\bQ}{\ve{Q}}

\newcommand{\bI}{\ve{I}}
\newcommand{\bL}{\ve{L}}

\newcommand{\tn}{\tilde{n}}

\newcommand{\dd}{\text{d}}
\newcommand{\ddd}{\dd R \dd \dot{R}}

\newcommand{\Rdot}{\dot{R}}
\newcommand{\Rddot}{\ddot{R}}

\newcommand{\brho}{\boldsymbol{\rho}}
\newcommand{\bLambda}{\boldsymbol{\Lambda}}

\newcommand{\eps}{\varepsilon}

\newcommand{\vbF}{\vec{\ve{F}}}
\newcommand{\vbg}{\vec{\ve{g}}}
\newcommand{\vbq}{\vec{\ve{q}}}
\newcommand{\vbr}{\vec{\ve{r}}}

\newcommand{\hx}{\hat{x}}

\newcommand{\hbq}{\widehat{\ve{q}}}
\newcommand{\hbg}{\widehat{\ve{g}}}
\newcommand{\hw}{\widehat{w}}

\newcommand{\hR}{\widehat{R}}
\newcommand{\hRdot}{\widehat{\dot{R}}}

\newcommand{\trho}{\widetilde{\rho}}
\newcommand{\tbrho}{\widetilde{\boldsymbol{\rho}}}
\newcommand{\tx}{\widetilde{x}}
\newcommand{\tbx}{\widetilde{\ve{x}}}

\newcommand{\mom}{\mu}
\newcommand{\bmom}{\boldsymbol{\mu}}
\newcommand{\vbmom}{\vec{\bmom}}
\newcommand{\hbmom}{\widehat{\bmom}}

\newcommand{\EV}{\mathbb{E}}

\newcommand\Rey{\mbox{Re}}

\newcommand\Web{\mbox{We}}

\makeatletter
\AtBeginDocument{\let\LS@rot\@undefined}
\makeatother

\newcommand{\overbar}[1]{\mkern 1.5mu\overline{\mkern-1.5mu#1\mkern-1.5mu}\mkern 1.5mu}

\usepackage{epsf}
\usepackage{stmaryrd}
\usepackage{setspace}
\usepackage{enumerate}
\usepackage{footnote}
\usepackage{scalefnt}
\usepackage{xfrac}
\usepackage{transparent}
\usepackage{rotate}
\usepackage{array}
\usepackage[normalsize]{subfigure}
\usepackage{amsmath}
\usepackage{amssymb}
\usepackage{graphicx}
\usepackage{float}
\usepackage{microtype}
\usepackage{siunitx}
\usepackage{xcolor}
\usepackage[margin=1in]{geometry}
\usepackage{listings}
\usepackage{arydshln}
\usepackage{lmodern}
\usepackage{upquote}
\usepackage{algorithm}
\usepackage{algpseudocode}
\usepackage{enumitem}
\usepackage[title]{appendix}

\newcommand\algostretch{1.35}

\usepackage[labelfont=bf,justification=raggedright,singlelinecheck=false]{caption}
\begin{document}

\begin{frontmatter}

\title{%
{\large\bfseries Conditional moment methods for polydisperse cavitating flows}}

\author[1]{Spencer H.\ Bryngelson}
\ead{shb@gatech.edu}
\author[2,3]{Rodney O.\ Fox}
\author[4]{Tim Colonius}

\address[1]{School of Computational Science \& Engineering, 
Georgia Institute of Technology, 
Atlanta, GA 30332, USA}

\address[2]{Department of Chemical and Biological Engineering, 
Iowa State University, 
Ames, IA 50011, USA}

\address[3]{Center for Multiphase Flow Research and Education, 
Iowa State University, 
Ames, IA 50011, USA}

\address[4]{Department of Mechanical and Civil Engineering, 
California Institute of Technology, 
Pasadena, CA 91125, USA}

\date{\today}

\begin{abstract}
    The dynamics of cavitation bubbles are important in many flows, but their  small sizes and high number densities often preclude direct numerical simulation.
    We present a computational model that averages their effect on the flow over larger spatiotemporal scales.
    The model is based on solving a generalized population balance equation (PBE) for nonlinear bubble dynamics and explicitly represents the evolving probability density of bubble radii and radial velocities. 
    Conditional quadrature-based moment methods (QBMMs) are adapted to solve this PBE.
    A one-way-coupled bubble dynamics problem demonstrates the efficacy of different QBMMs for the evolving bubble statistics. Results show that enforcing hyperbolicity during moment inversion (CHyQMOM) provides comparable model-form accuracy to the traditional conditional method of moments and decreases computational costs by about ten times for a broad range of test cases.
    The CHyQMOM-based computational model is implemented in MFC, an open-source multi-phase and high-order-accurate flow solver.
    We assess the effect of the model and its parameters on a two-way coupled bubble screen flow problem.
\end{abstract}


\end{frontmatter}

\section{Introduction}

Cavitating flows arise near ship propellers~\citep{cook28}, in hydraulic machinery~\citep{arndt81}, over spillways~\citep{falvey90}, and during medical therapies like lithotripsy and histotripsy~\citep{cleveland00,pishchalnikov03,maxwell11}.  In these flows, polydisperse bubble clouds are generated directly through nucleation or break-up and shedding larger vapor regions.
Directly simulating the flow of these dispersions is challenging since the associated bubble dynamics frequently occur at smaller length and time scales than the background flow~\citep{brennen95}.

In the dilute limit, a strategy for overcoming this scale separation recognizes that the dynamics of each bubble are unimportant compared to statistics of the bubble population.
Ensemble phase-averaging~\citep{zhang94,zhang94b}, for example, results in equations for the continuous liquid face forced, in a two-way-coupled manner, by statistics of (typically) the bubble radius radial velocity.  The associated bubble variables become stochastic, Eulerian fields.  This contrasts against Lagrangian models that track, model, and average over samples of individual particles~\citep{bryngelson19,capecelatro13}.  Unless the bubble populations are such that any small volume (compared to the mixture-averaged flow field variations) contains a sufficient number of bubbles to describe the statistics faithfully, these models simulate at best a single realization of a random process.

The population balance equation (PBE) represents the evolution of a dispersion's probability (or number density function) according to a set of internal variables~\citep{randolph87,ramkrishna00}.  
The internal variables must provide sufficient information to close each particle's dynamic and thermodynamic evolution. 
The PBE is commonly used to model the particle size distribution. It depends solely on particle motion, coagulation, and breakup~\citep{fox03,rigopoulos10,lee15}, wherein the internal variables are the spatial coordinates and velocities.
PBE-based models have successfully modeled flowing soot during combustion processes~\citep{Mueller2009}, aerosol sprays~\citep{sibra2017}, and more.
For oscillating bubbles, the internal variables must include the bubble radius, bubble radial velocity, and sufficient information regarding the bubble contents, which can often characterize with a single equilibrium radius~\citep{brennen95}. 
When appropriate, these three internal variables can be appended to those associated with relative motion, coalescence, and break up~\citep{carrica99,heylmun19}.

The current work formulates a PBE-based model for the distribution of bubble radius, radial velocity, and equilibrium radius. 
Once formulated, the PBE is a PDE in independent coordinates associated with each internal variable (as well as spatiotemporal coordinates associated with the continuous flow in which the bubbles reside).  
Solving this  six-dimensional PDE via the method of lines is intractable.
Instead, the method of moments represents and evolves the number density function via its statistical moments~\citep{hulburt64}.
Other approaches to solving PBEs exist, like class~\citep{hounslow88,vanni00}, particle~\citep{patterson11}, and Monte Carlo~\citep{zhao07,rosner03} methods, though these are comparatively inefficient for multivariate PBEs or large simulations~\citep{zucca07}.

The ensemble phase-averaging methods determine the governing flow equations for oscillating bubble populations. 
As described in section~\ref{s:model}, we develop an Euler--Euler approach to represent the flow and the averaged bubble dynamics at the sub-grid level~\citep{ando11,bryngelson19}. 
This method is two-way-coupled between the dispersed bubbles and suspending liquid.
In particular, moments of the oscillating bubble dynamics alter the effective pressure and void fraction evolution equation. 

Next, in section~\ref{s:pbe}, we develop and verify the conditional moment methods discussed above to close these equations.  
This requires a dynamical/thermodynamical model for each bubble. 
Even under the assumption of spherical bubbles, the most general form of such a model consists of a set of PDEs for the balance of mass, momentum, energy. 
Under further assumptions, however, these can be reduced to a set of ODEs for the bubble radius and radial velocity, with equilibrium radius as a parameter~\citep{plesset77}. 
The set of ODEs for each particle determines the derivatives of the internal variables that close the moment transport equations~\citep{frenklach87}. 
Quadrature-based moment methods then invert the moment set for a quadrature rule approximating the quantities required to close the moment transport equations~\citep{mcgraw97,marchisio13}.
With multiple independent variables, conditional quadrature moment methods are computationally preferable~\citep{yuan11}.
This work implements the conditional quadrature method of moments (CQMOM~\citep{yuan11}) and its hyperbolically constrained version, CHyQMOM~\citep{Fox2018,patel19,fox2021hyperbolic}.

Section~\ref{s:numerics} describes our interface-capturing numerical algorithm 
for solving the resulting system of equations. 
We demonstrate the methodology in section~\ref{s:coupled} by considering an acoustically excited bubble screen.
Lastly, section~\ref{s:conclusion} summarizes our results and potential next steps for extending and analyzing the PBE-based method.

\section{Model formulation}\label{s:model}

\subsection{Compressible flow equations}\label{s:floweq}

A dilute suspension of dynamically evolving bubbles flow in a compressible liquid is considered.  For simplicity, we assume no slip between the bubbles and the surrounding liquid and that the gas density is much smaller than the liquid density.
Under these assumptions, the mixture-averaged form of the compressible flow equations are
\begin{align}
    \frac{ \partial \rho }{\partial t} + \nabla \cdot ( {\rho \bu} ) =& \, 0, \nonumber \\
    \frac{ \partial {\rho \bu} }{\partial t} + \nabla \cdot ( \rho \bu \bu + p \bI ) =& \, 0, \label{e:euler}  \\
	\frac{ \partial {E} }{\partial t} + \nabla \cdot ({E} + p) \bu  =& \, 0, \nonumber
\end{align}
where $\rho$, $\bu$, $p$, and $E$ are the density, velocity vector, pressure, and total energy. 
Terms associated with the bubbles modify these quantities and transport them in space according to ensemble phase-averaging, discussed next.

\subsection{Ensemble phase-averaging}\label{s:avg}

The ensemble-averaged equations follow from~\citet{zhang94} and~\citet{bryngelson19}.
The disperse phase has a void fraction $\alpha$ and is assumed to be a dilute ($\alpha \ll 1$) population of spherical bubbles.
The bubbles are represented statistically via random variables $R$, $\Rdot$, and $R_o$ corresponding to the instantaneous bubble radius, time derivative, and equilibrium bubble radius.
Section~\ref{s:bub} presents this bubble dynamics model in detail.
The mixture-averaged pressure field is computed as 
\begin{gather}
    p(\bx,t) = (1-\alpha) p_\ell +
	\alpha  \left(
		\frac{\overbar{R^3 p_{bw} }}{\overbar{R^3}} - \rho \frac{ \overbar{ R^3 \dot{R}^2 }}{ \overbar{R^3} }
	\right),
    \label{e:pressure}
\end{gather}
where $p_{bw}$ are the associated bubble wall pressure and $p_\ell(\bx,t)$ is the liquid pressure according to the stiffened-gas equation of state~\citep{menikoff89}:
\begin{gather}
	\Gamma_\ell p_\ell + \Pi_{\infty,\ell} = 
    \frac{1}{1-\alpha} \left( E - \frac{1}{2} \rho \bu^2 \right).
	\label{e:SEOS}
\end{gather}
The coefficients of~\eqref{e:SEOS} represent water, with specific heat ratio $\gamma_\ell  = 1 + 1/\Gamma_\ell = 7.15$ and stiffness $\Pi_{\infty,\ell} = \SI{356}{\mega\pascal}$~\citep{maeda18}.

The bubble number density per unit volume $n(\bx,t)$ is conserved in the absence of coalescence or breakup:
\begin{gather}
    \frac{\partial n }{\partial t } + \nabla \cdot ( n \bu ) = 0.
    \label{e:consn}
\end{gather}
For the spherical bubbles considered here, $n$ is related to the void fraction $\alpha$ via 
\begin{gather}
    \alpha(\bx,t) = \frac{4}{3} \pi { \overbar{R^3} } n(\bx,t),
    \label{e:ndf}
\end{gather}
and thus the void fraction $\alpha(\bx,t)$ transports as
\begin{gather}
    \frac{\partial \alpha }{\partial t } + \bu \cdot \nabla \alpha =
	3 \alpha \frac{ \overbar{R^2 \dot{R} }}{ \overbar{R^3} },
    \label{e:alpha}
\end{gather}
where the right-hand-side represents the change in void fraction due to bubble growth and collapse.

The over-barred terms appearing in~\eqref{e:pressure}, \eqref{e:ndf}, and \eqref{e:alpha},
\begin{gather}
    \overbar{ R^3 \dot{R}^2},  \;
    \overbar{R^3}, \;
    \overbar{R^2 \dot{R} }, \; \text{and }
    \overbar{R^3 p_{bw}}.
    \label{e:fullmoments}
\end{gather}
denote average quantities of the bubble dispersion. 
In particular, they are raw moments $\mom_{lmn}$ with respect to a bubble number density function $f(R,\Rdot,R_o)$,
\begin{gather}
    \mom_{lmn} = 
    \overbar{R^l \Rdot^m R_o^n} = 
    \int_\Omega R^l \Rdot^m R_o^n f(R,\Rdot,R_o) \, \ddd \dd R_o,
    \label{e:raw}
\end{gather}
which are computed via the methods of section~\ref{s:pbe}.

\subsection{Bubble dynamics model}\label{s:bub}

Even for spherical models, dozens of available models employ differing assumptions about the bubble contents and simplifications of the physics. 
These models range from a system of two or more ODEs up to a set of PDEs (in a radial coordinate and time) describing the balance of mass, momentum, and energy of each bubble. 
PDE-based models are deemed intractable (at present), and we focus on ODE-based modeling.  
We choose one of the simplest possible models to demonstrate our methodology but note that our framework can be extended to arbitrarily complex ODE-based models.

We assume the spherical bubbles are filled with noncondensible gas and that, insofar as their dynamics are concerned, the liquid may be assumed incompressible. 
We assume that the gas undergoes a polytropic process during compression and expansion. 
Under these assumptions, the bubble radius is governed by a Rayleigh--Plesslet-like equation
\begin{gather}
    R \ddot{R} + \frac{3}{2} \Rdot^2 + \frac{4}{\Rey} \frac{\Rdot}{R} = 
    \left( \frac{R_o}{R} \right)^{3\gamma}  - 
    \frac{1}{C_p} -
    \frac{2}{\Web \, R_o} \left[ \frac{R_o}{R} - \left(\frac{R_o}{R}\right)^{3\gamma} \right],
    \label{e:rpe}
\end{gather}
which is dimensionless via the reference bubble size $R_o^\ast$ and ambient liquid pressure $p_0$, and density $\rho_0$.  
The polytropic index is $\gamma$; in examples below we use $\gamma=1.4$. 
In~\eqref{e:rpe}, $C_p \equiv p_0/p_\ell$ is the forcing pressure ratio and Reynolds and Weber numbers correspond to viscous and surface tension effects as
\begin{gather}
    \Rey \equiv \sqrt{\frac{p_0}{\rho_0}} \frac{R_o^\ast}{\nu_0}
    \quad \text{and} \quad 
    \Web \equiv \frac{p_0 R_o^\ast}{S},
\end{gather}
where $S$ is the air--water surface tension coefficient and $\nu_0$ is the liquid kinematic viscosity.   
Under these conditions, the bubble wall pressure of~\eqref{e:pressure} simplifies to $p_{bw} = \left(R_o/R\right)^{3\gamma}$ and the last moment of~\eqref{e:fullmoments} reduces to $\overbar{R^3  (R_o/R)^{3\gamma}}$.

\section{Population balance formulation}\label{s:pbe}

\begin{figure}
    \centering
    \includegraphics{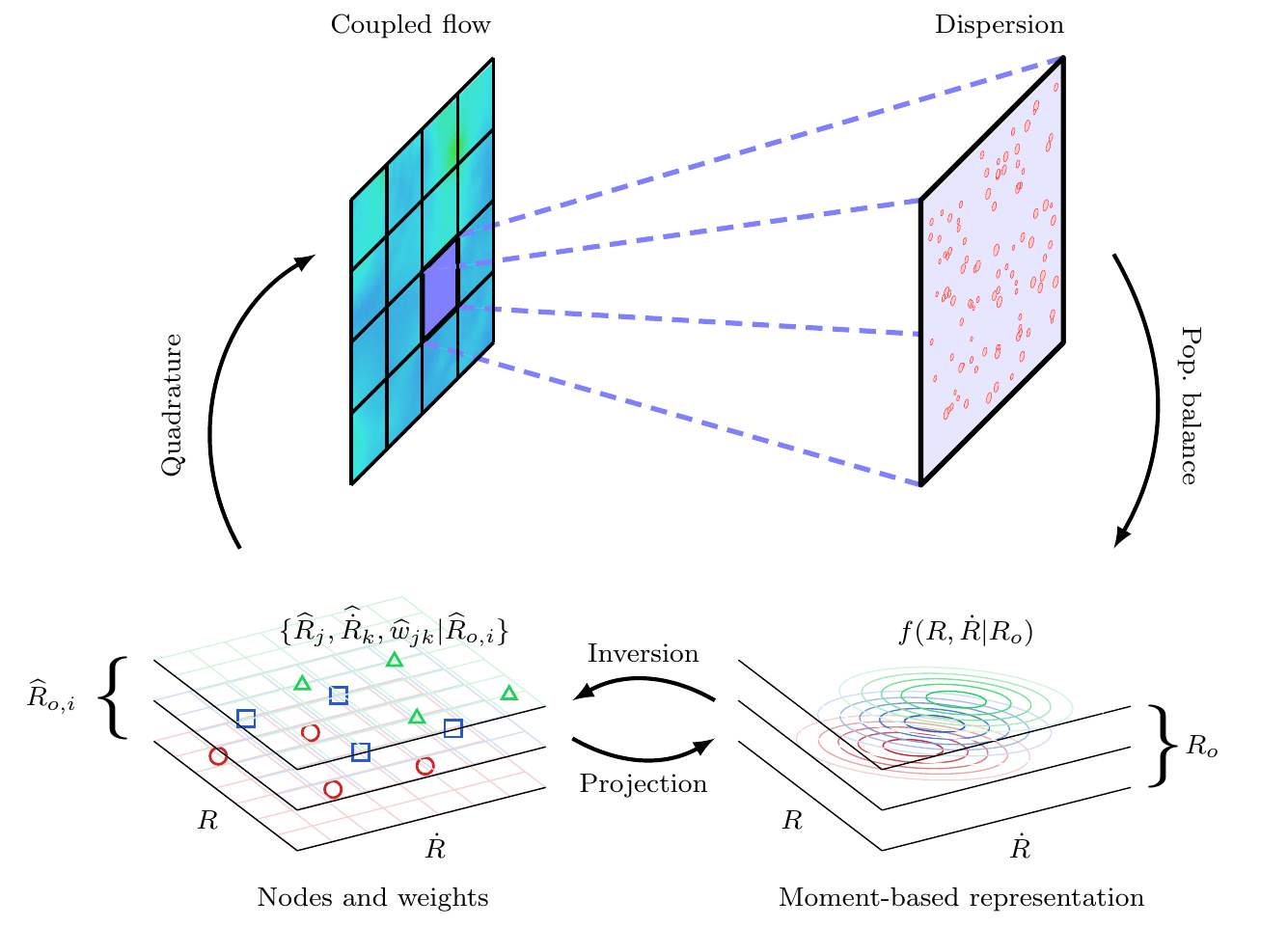}
    \caption{Illustration of the quadrature-based moment method for a fully coupled bubbly flow.}
    \label{f:model_schematic}
\end{figure}

The population balance equation (PBE) 
\begin{gather}
    \frac{ \partial f}{\partial t} + 
	\frac{\partial}{\partial R} (f \Rdot ) + 
	\frac{\partial}{\partial \Rdot} (f \ddot{R} ) = 0
    \label{e:master}
\end{gather}
governs the conditional PDF $f(R,\dot{R}|R_o)$ in the absence of bubble coalescence or breakup, though this approach can naturally accommodate these effects if required.
Figure~\ref{f:model_schematic} summarizes the PBE--quadrature approach.

\subsection{Method of moments}\label{s:mom}

Following the usual procedure, a finite set of raw moments $\vbmom$ represent $f$ per~\eqref{e:raw}~\citep{fox03}.
The specific moments that make up $\vbmom$ depend on the inversion algorithm and are defined in the appendices~A and~B.
These moments transport on the grid and evolve according to the bubble dynamics of section~\ref{s:bub} as
\begin{gather}
    \frac{\partial n \vbmom}{\partial t} + \nabla \cdot (n \vbmom \bu) = 
    n {\dot\vbmom} = 
    n \vbg
	\label{e:transport}
\end{gather}
where 
\begin{gather}
    g_{lmn} = 
    l \mom_{l-1,m+1,n} + 
    m \iiint_\Omega \Rddot  R^l \Rdot^{m-1} R_o^n f(\vbmom) \, \ddd \dd R_o
	\label{e:rhs}
\end{gather}
and $\Omega = \Omega_R \times \Omega_{\Rdot} \times \Omega_{R_o} = (0,\infty) \times (-\infty,\infty) \times (0,\infty)$.
The integrand of~\eqref{e:rhs} is closed via the bubble dynamics model~\eqref{e:rpe} for $\Rddot$, and the integral is approximated via quadrature, which is discussed next.

\subsection{Conditional quadrature moment inversion}\label{s:cqmom}

Since $R_o$ is not a dynamic variable, the number density function is split as
\begin{gather}
    f(R,\Rdot,R_o) = f(R,\Rdot|R_o) f(R_o).
\end{gather}
The raw moments~\eqref{e:raw} are then
\begin{align}
    \mom_{lmn} 
        &\equiv \int_{\Omega_{R_o}} f(R_o) R_o^m \mom_{lm}(R_o) \, \dd R_o \\
        &\approx \sum_{i=1}^{N_{R_o}} w_i \hR_{o,i}^n \, \mom_{lm}(\hR_{o,i}),
    \label{e:full}
\end{align}
with $N_{R_o}$ time-independent weights $w_i$ and nodes $\hR_{o,i}$.
Simpson's rule computes these nodes and weights and the accuracy in the approximation of~\eqref{e:full} depends on $N_{R_o}$.
Other numerical integration methods are suitable and will be discussed in section~\ref{s:closurerr}.
The $R_o$-conditioned moments are
\begin{align}
    \mom_{lm}(\hR_{o,i}) 
        &\equiv \iint_{\Omega_{R,\Rdot}} f(R,\Rdot|\hR_{o,i}) R^l \Rdot^m \, \dd R \dd \Rdot \\
        &\approx \sum_{j=1}^{N_R} \sum_{k=1}^{N_{\Rdot}} \left[ \hw_{j,k} \hR_j^l \hRdot_k^m \right]_{\hR_{o,i}}.
    \label{e:conditioned}
\end{align}
The moment indices comprising the moment set of~\eqref{e:conditioned} are determined by the conditional quadrature moment method used to invert those moments for a set of quadrature points and weights (and the number of points desired)~\citep{yuan11,patel19}.
In particular, $\mom_{lm}(\hR_{o,i})$ is traded for quadrature points $\{ \hR_j, \hRdot_k \}(\hR_{o,i})$ and weights $\hw_{j,k}(\hR_{o,i})$ for each $i = 1,\dots,N_{R_o}$ (with $j = 1,\dots,N_{R}$; $k = 1,\dots,N_{\Rdot}$).
The total moments~\eqref{e:full} are then approximated by substituting~\eqref{e:conditioned} into \eqref{e:full} as 
\begin{gather}
    \mom_{lmn} = 
        \sum_{i=1}^{N_{R_o}} 
        w_i \hR_{o,i}^n 
        \sum_{j=1}^{N_R} \sum_{k=1}^{N_{\Rdot}} \left[ \hw_{j,k} \hR_j^l \hRdot_k^m \right]_{\hR_{o,i}}.
\end{gather}
The CHyQMOM and CQMOM algorithms are described in appendices~\ref{a:chyqmom} and~\ref{a:cqmom}. 
Their implementation is verified by ensuring that the error in closing a linear set of moment transport equations is comparable to a finite-precision round-off~\citep{bryngelson20_qbmm}.

\subsection{Model performance}\label{s:uncoupled}

We consider the ability of the QBMMs to represent the statistics of the bubble dynamics.
For this example, we take a monodisperse population with the same equilibrium radius $R_o = 1$ (and thus $N_{R_o} = 1$).
$R$ and $\Rdot$ are initialized with independent log-normal and normal distributions with variances $\sigma_R^2$ and $\sigma_{\Rdot}^2$, respectively. The bubbles are forced by a step-change in pressure at $t=0^+$ from equilibrium to a value $C_p$ (which is varied).  This represents a stringent test of model fidelity as high values of $C_p$ will induce strongly nonlinear bubble dynamics.

\begin{figure}
    \centering
    \includegraphics{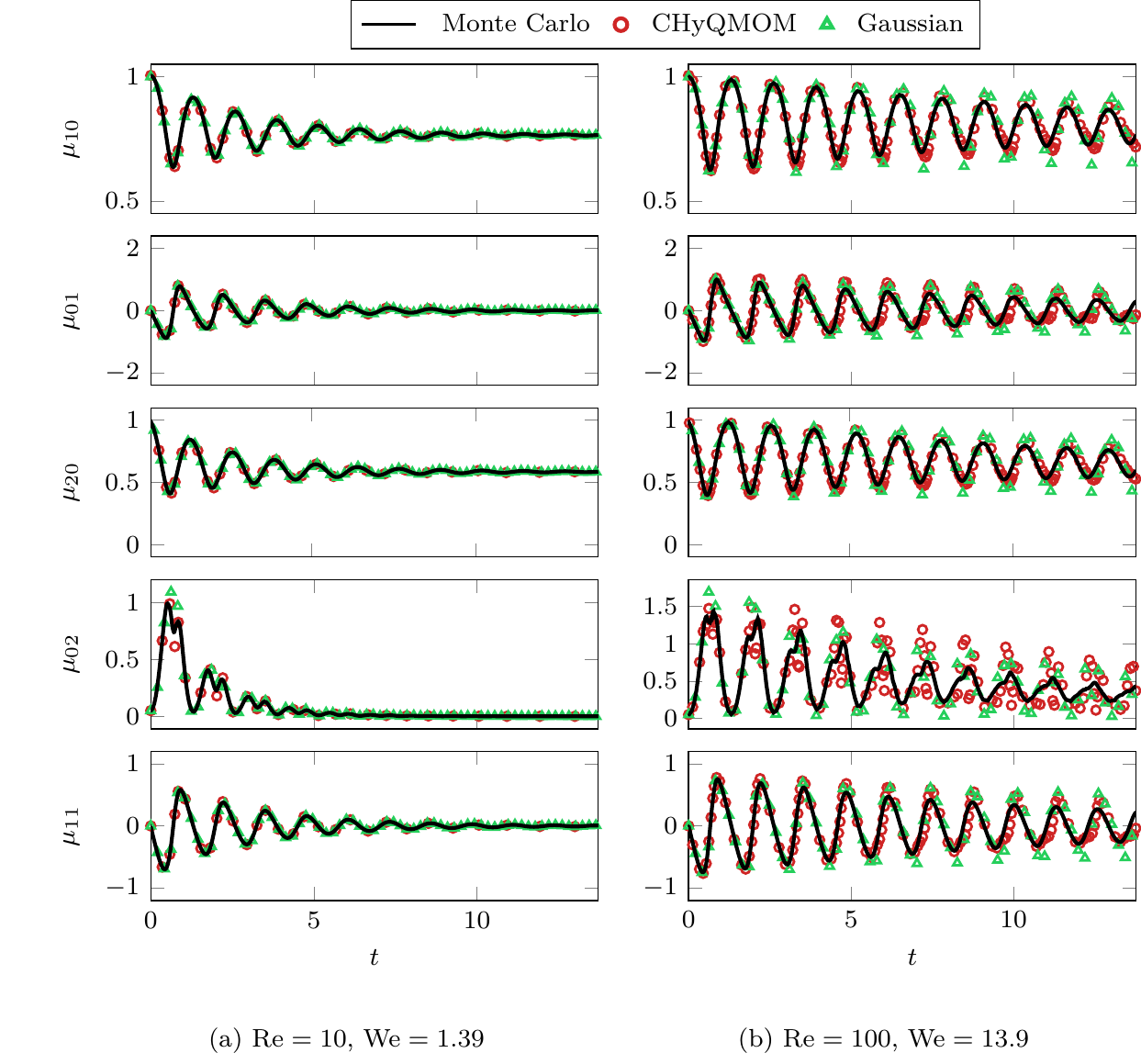}
    \caption{
        Example first- and second-order moments $\mom$ as labeled for bubbles with Reynolds and Weber numbers corresponding to (a) $R_o^\ast = \SI{1}{\micro\meter}$ and (b) $R_o^\ast = \SI{10}{\micro\meter}$.
        The dimensionless initial shape parameters are $\sigma_R = \sigma_{\Rdot} = 0.2$.
        The CHyQMOM implementation uses two quadrature nodes in each internal coordinate direction ($2 \times 2$), and ``Gaussian'' corresponds to Gaussian closure~\citep{bryngelson19}.
        Monte Carlo serves as an exact solution.
        The final time is $T = 13.9$ and represents 10 natural periods of the largest bubble.
    }
    \label{f:nonlinmoments}
\end{figure}

Figure~\ref{f:nonlinmoments} shows the evolution of the carried moment set for $C_p = 0.3$.  To assess the accuracy, the CHyQMOM results are compared against Monte-Carlo solutions that used $10^4$ samples to ensure that the sampling error was at least 10-times smaller than the QBMM model-form error.  Thus the Monte-Carlo solutions can be regarded as exact solutions for these comparisons.
The radial moments $\mom_{10}$ and $\mom_{20}$ show how the bubbles oscillate in response to the change in external pressure, eventually reaching a statistical equilibrium~\citep{colonius08}.
These oscillations are damped rapidly for smaller bubbles due to stronger viscous effects (smaller $\Rey$).
Thus, larger bubbles entail larger model-form errors as they build over time.
This is most prominent for the $\Rdot$ moments, such as $\mom_{02}$, and smaller bubbles ((b) $R_o^\ast = \SI{10}{\micro\meter}$) as the velocity statistics change more quickly than the radial ones.

\begin{figure}
    \centering
    \includegraphics{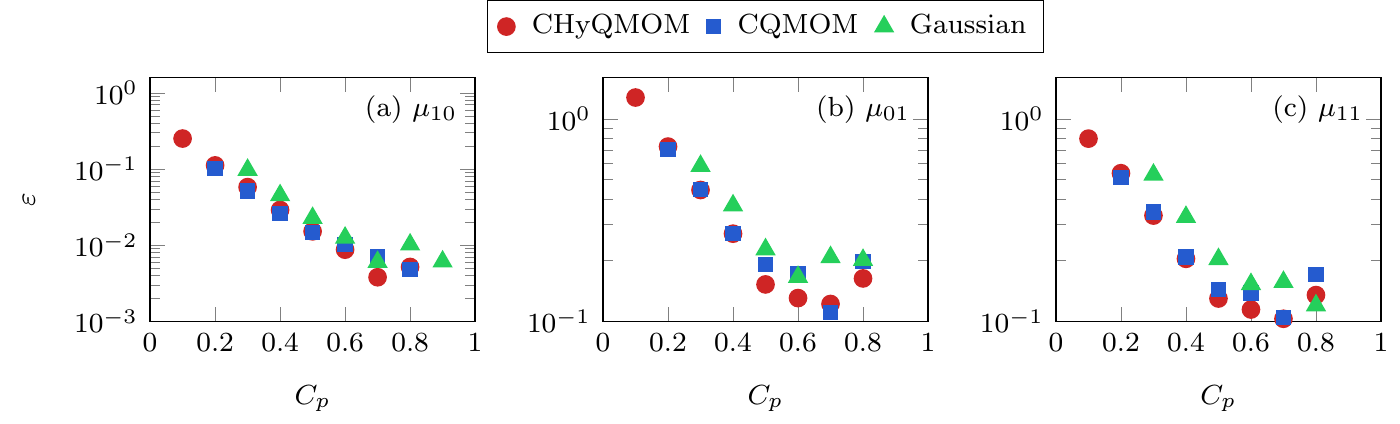}
    \caption{Relative model-form errors for varying pressure ratio $C_p$ and closure methods as labeled. All cases have $\Rey = 100$ and $\Web = 13.9$.}
    \label{f:errors}
\end{figure}

A QBMM model-form relative $L_2$ error is
\begin{gather}
    \eps \equiv 
    \frac{1}{N_t} \sqrt{ 
        \sum_{i=1}^{N_t} 
        \left[
        \frac{\mom^\mathrm{(QBMM)}(t_i) - \mom^\mathrm{(MC)}(t_i)}{\mom^\mathrm{(MC)}(t_i)}
        \right]^2
    },
    \label{e:error}
\end{gather}
where Monte-Carlo (MC) simulations serve as surrogate truth data and $t_i$ are $N_t = 10^3$ uniformly spaced times in the time interval $t \in [0,T]$.
Figure~\ref{f:errors} shows $\eps$ for select first- and second-order moments $\mom$.
We see that $\eps$ is smaller for CHyQMOM and CQMOM than Gaussian closure, though the difference is modest and more strongly associated with $C_p$.
The larger errors for smaller $C_p$ are associated with stronger bubble dynamics and the formation of non-Gaussian statistics, like skewness and kurtosis, that these closures do not represent~\citep{Bryngelson2020}.
One can represent higher-order statistics by carrying higher-order moments and thus inverting for a higher-order quadrature rule.
However, this is computationally cumbersome and numerically unstable in the cases tested here.

\begin{figure}
    \centering
    \includegraphics{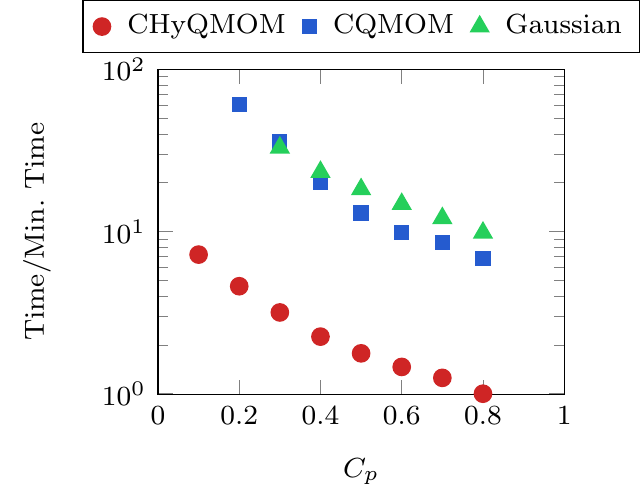}
    \caption{Total simulation times for the cases of figure~\ref{f:errors}.
    A single-core of a $\SI{2.8}{\giga\hertz}$~Intel~i7-7700HQ performed the computations.
    Only relative times are relevant, so the cheapest simulation normalizes the scale (CHyQMOM, $C_p = 0.8$).
    }
    \label{f:timing}
\end{figure}

Model accuracy was similar for all of the closure methods, but the computational time to solution is not. 
Figure~\ref{f:timing} shows the total relative simulation time for the cases of figure~\ref{f:errors} under the same adaptive time-step tolerance.
CHyQMOM simulations are about 10-times cheaper than those using CQMOM or Gaussian closures, even though the accuracy is the same.
This is attributed to larger time step sizes (given the same error constraint) and a smaller carried moment set than CQMOM.
Due to its low cost for the same accuracy, the fully-coupled simulation algorithm employs CHyQMOM and is discussed next.

\section{Interface-capturing flow solution algorithm}\label{s:numerics}

The QBMM approach of the previous section is solved using interface-capturing numerics.
MFC, an open-source flow solver, is the basis for the implementation~\citep{bryngelson19_CPC}.
A brief description of the algorithm follows.

The governing equations~\eqref{e:euler}, \eqref{e:alpha}, and \eqref{e:transport} combine as
\begin{gather}
    \frac{\partial\vbq_c}{\partial t} + \nabla \cdot \vbF = \vbr
    \label{e:governing}
\end{gather}
where $\vbq_c \equiv \{ \rho, \rho \bu, E, \alpha, n \vbmom \}$ are the conservative variables, $\vbF$ are the advective fluxes, and $\vbr$ are diffusive source terms.
Finite volumes with uniform size $\Delta$ discretize the domain.

\subsection{Flow state initialization}

We initialize the flow as follows.
The number density function $f(R,\Rdot,R_o)$ is independently distributed with log-normal, normal, and log-normal shapes in the $R$, $\Rdot$, and $R_o$ directions with expected values $\EV[R] = \EV[R_o] = 1$ and $\EV[\Rdot] = 0$ and shape parameters $\sigma_\cdot$. 

The NDF $f$ initializes the moment set $\vbmom$ via integration.
The primitive variables $\vbq_p \equiv \{ \rho, \bu, p, \alpha, \vbmom \}$ are initialized as appropriate.
The bubble number density $n$ follows from~\eqref{e:ndf}, $\vbq_p$, and $\mom_{300}$.
Lastly, the known primitive variables are used to compute the conservative ones as $\vbq_c \equiv \{ \rho, \rho \bu, E, \alpha, n \vbmom \}$.

\begin{algorithm}
    \setstretch{\algostretch}
    \begin{algorithmic}[1]
        \State $f(R,\Rdot,R_o) \leftarrow$ Presumed form, uncorrelated, $\{\mom_R,\mom_{\Rdot},\mom_{R_o}\}$ and $\{\sigma_R,\sigma_{\Rdot},\sigma_{R_o}\}$
        \State $\vbmom \leftarrow f$ and \eqref{e:raw}
        \State $\vbq_p \equiv \{ \rho, \bu, p, \alpha, \vbmom \} \leftarrow$ Patches
        \State  $n \leftarrow \alpha, \mom_{300}$ and \eqref{e:ndf}
        \State  $\vbq_c \leftarrow n, \vbq_p$
    \caption{Flow initialization procedure}\label{a:init}
    \end{algorithmic}
\end{algorithm}

\subsection{Flux divergence computation}

A fifth-order-accurate WENO~\citep{jiang1996efficient} scheme reconstructs the primitive variables $\vbq_p$ and the HLLC approximate Riemann solver~\citep{toro94} computes the fluxes.
Algorithm~\ref{a:base} describes this process in detail.
High-order WENO reconstructions do not guarantee that the reconstructed moments are realizable, though the moment sets remained invertible in the subsequent simulations.

\begin{algorithm}
    \setstretch{\algostretch}
    \begin{algorithmic}[1]
        \State $\vbq_p \leftarrow \vbq_c$ and~\eqref{e:SEOS}
        \State $\hbq_p \leftarrow \texttt{WENO}(\vbq_p)$  
        \State $\{\hR,\hRdot,\hw\}_{jk} \leftarrow \texttt{QBMM}(\hbmom)$
        \State  $\hbg, \{ 
                \overbar{ R^3 \dot{R}^2}, 
                \overbar{R^3}, 
                \overbar{R^2 \dot{R}},
                \overbar{R^3 p_{bw}}
        \} \leftarrow \{\hR_{o},w\}_{i}, \{\hR,\hRdot,\hw\}_{jk,i}$
        \State $\vbF \leftarrow 
                \texttt{HLLC}(
                    \hbq_p, \hbg, \overbar{\{\cdot\}}
                    )$
        \State $\nabla \cdot \vbF \leftarrow \vbF$
    \caption{Algorithm for computing the flux divergence.}
    \label{a:base}
    \end{algorithmic}
\end{algorithm}

\subsection{Time stepping}

The conservative variables are integrated in time using third-order-accurate SSP--RK3 time integration~\citep{Gottlieb2001}.
Once the spatial derivatives have been approximated, \eqref{e:governing}~becomes a semi-discrete system of ordinary differential equations in time. 
We treat the temporal derivative using a Runge--Kutta time-marching scheme for the state variables. 
To achieve high-order accuracy and avoid spurious oscillations, we use the third-order-accurate total variation diminishing scheme of~\citet{gottlieb98}:
\begin{align}
	\vbq_{c}^{\,(1)} &= \vbq_{c}^{\,\tn} + 
            \Delta t \bL(\vbq_{c}^{\,\tn}), \nonumber\\
	\vbq_{c}^{\,(2)} &= \frac{3}{4} \vbq_{c}^{\,\tn} + 
            \frac{1}{4} \vbq_{c}^{(1)} + 
            \frac{1}{4} \Delta t \bL(\vbq_{c}^{\,(1)}), \label{e:time} \\
	\vbq_{c}^{\,\tn+1} &= \frac{1}{3} \vbq_{c}^{\,\tn} + 
            \frac{2}{3} \vbq_{c}^{\,(2)} + 
            \frac{2}{3} \Delta t \bL(\vbq_{c}^{\,(2)}), \nonumber
\end{align}
where superscripts~$(1)$ and $(2)$ indicate intermediate time-step stages, $\bL$ represents the portion of~\eqref{e:governing} that does not include the time derivative, and $\tilde{n}$ is the time-step index.

\section{Application to polydisperse bubble screens}\label{s:coupled}

\subsection{Problem setup}

\begin{figure}
    \centering
    \includegraphics{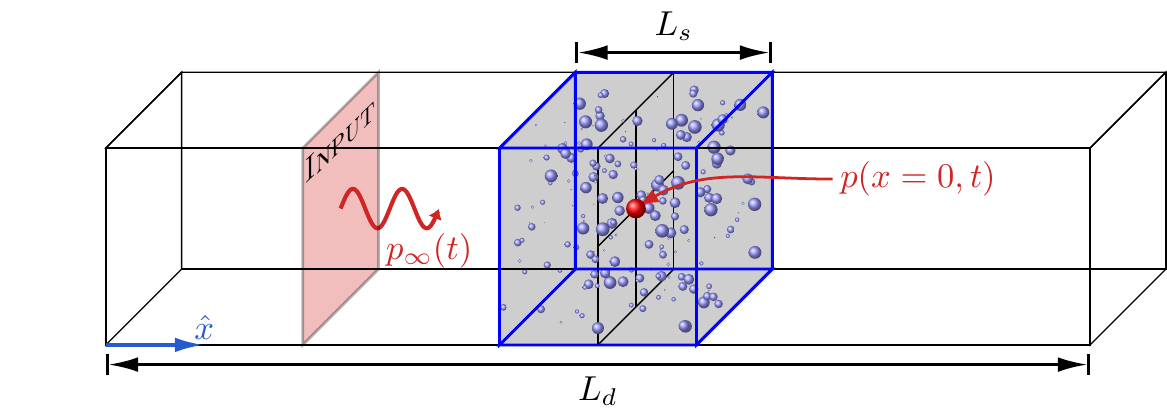}
    \caption{
        Schematic of the acoustically excited bubble screen test problem.
        The screen is centered at $x=0$.
    }
    \label{f:bubble_screen_setup}
\end{figure}

We assess the statistics of bubble dynamics in an idealized model problem consisting of an acoustically excited dilute bubble screen.
The bubble screen parameterization matches that of~\citet{bryngelson19}, with a length of $L_s = \SI{5}{\milli\meter}$, initial void fraction $\alpha_o = 10^{-4}$, median bubble equilibrium size $R_o^\ast = \SI{10}{\micro\meter}$ and log-normal variance $\sigma_{R_o}$.
The domain is $L_d = 5 L_s$ long, and its boundaries are non-reflective via characteristic-based boundary conditions~\citep{thompson86}.
The one-way (positive $\hx$-direction) sound wave $p_\infty(t)$ is generated via source terms in the governing equations~\eqref{e:governing} according to~\citet{bryngelson19}.
Its form is a single period of a sinusoid with peak amplitude $0.3 p_0$ and frequency $\SI{300}{\kilo\hertz}$.

\subsection{Bubble screen behavior}

\begin{figure}
    \centering
    \includegraphics{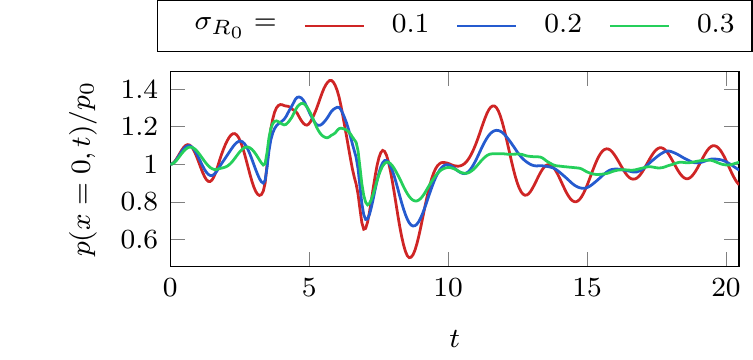}
    \caption{Bubble-screen-centered pressure for varying $R_o$ log-normal distributions with shape parameter $\sigma_{R_o}$ and fixed $\sigma_R = \sigma_{\Rdot} = 0.2$.}
    \label{f:probeR0}
\end{figure}

We start by considering a screen with fixed dynamic coordinate distributions $\sigma_R = \sigma_{\Rdot}$, but varying distributions of equilibrium sizes $\sigma_{R_o}$.
Polydispersity in $R_o$ is integrated via Simpson's rule $61$ quadrature points for all cases.
Figure~\ref{f:probeR0} shows the bubble screen pressure for these cases as they evolve in time.
For larger $\sigma_{R_o}$ (or broader distributions or bubble equilibrium sizes), the pressure is less oscillatory in time.
A similar observation was made by~\citet{bryngelson19} for cases with no $R$ or $\Rdot$ distributions.
In the case of figure~\ref{f:probeR0}, we instead observe high-frequency oscillations in addition to the long-wavelength behaviors associated with the impinging pressure wave $p_\infty$.
The origin of these oscillations is discussed next.

\begin{figure}
    \centering
    \includegraphics{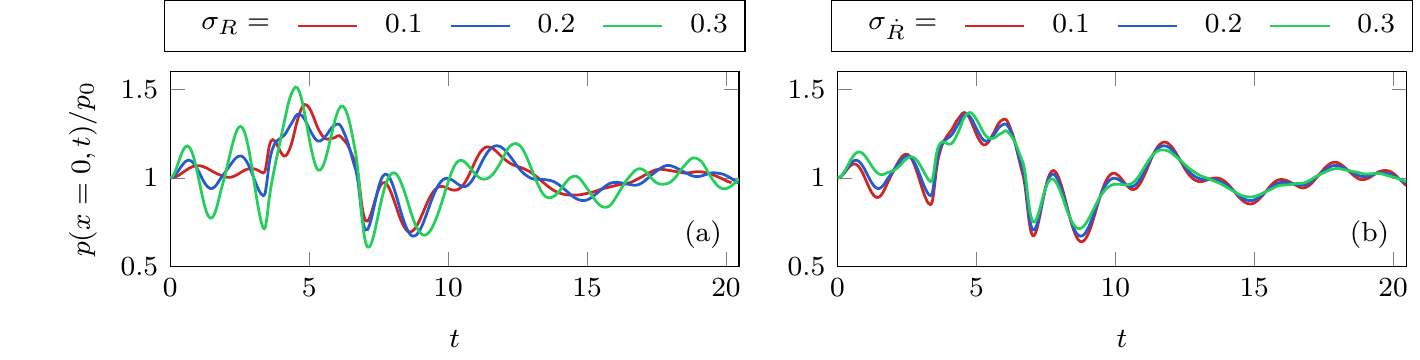}
    \caption{Bubble-screen-centered pressure before, during, and after excitement due to an acoustic wave. 
        The bubbles are polydisperse with log-normal $R_o$ distribution ($\sigma_{R_o} = 0.2$) and $\Rey = 10^3$.
        Variations in (a)~$\sigma_R$ and (b)~$\sigma_{\Rdot}$ are shown about a $\sigma_R = \sigma_{\Rdot} = 0.2$ representative state.
    }
    \label{f:probeRV}
\end{figure}

Figure~\ref{f:probeRV} shows the dynamics associated with a bubble screen in varying degrees of statistical disequilibrium, represented via different $\sigma_R$ and $\sigma_{\Rdot}$.
Figure~\ref{f:probeRV}~(a) fixes $\sigma_{\Rdot}$ and varies $\sigma_R$.
We observe the shorter-wavelength oscillatory behavior, observed in figure~\ref{f:probeR0}, becoming more prominent for larger $\sigma_R$.
These wavelengths are commensurate with the mean bubble natural frequencies, which superimpose the longer wavelength acoustics associated with the impinging $p_\infty$ wave.
Figure~\ref{f:probeRV}~(b) shows a smoother pressure profile for larger $\sigma_{\Rdot}$.
Phase-cancellation between the larger waves associated with broader $\sigma_{\Rdot}$ distributions and those of the $\sigma_R$ distributions may account for this behavior.
Notably, these behaviors are qualitatively similar to those associated with varying $R_o$ distribution widths.
Thus, parameterizing an $R_o$ distribution based on single-probe pressure measurements is insufficient.

\subsection{Closure errors}\label{s:closurerr}

We quantify the moment closure error, $\eps_c$, via the mismatch in bubble screen pressure $p(t,x=0)$ due to truncated $R_o$ integration.
Otherwise, the definition of $\eps_c$ has the same form as $\eps$ of~\eqref{e:error}, though the truth values are changed from Monte-Carlo simulations (which are unavailable) to high-resolution $N_{R_o} = 401$ simulations as
\begin{gather}
    \eps_c \equiv 
    \frac{1}{N_t} \sqrt{ 
        \sum_{i=1}^{N_t} 
        \left[
            \frac{p^\mathrm{(QBMM)}(t_i,x=0) - p^{(N_{R_o} = 401)}(t_i,x=0)}{p^{(N_{R_o} = 401)}(t_i,x=0)}
        \right]^2
    }.
    \label{e:errorc}
\end{gather}
This choice is made for two reasons.
First, extending the QBMM method to additional $R$ and $\Rdot$ quadrature points (or moments, equivalently) was found numerically unstable for most bubble cavitation problems.
One approach to addressing this specific problem is introducing a recurrent neural network to correct the quadrature points and weights~\citep{charalampopoulos2021hybrid}, though we do not discuss it further here.
Second, we will see that the closure errors are most strongly associated with $N_{R_o}$, and thus focus on the influence of this parameter.

\begin{figure}
    \centering
    \includegraphics{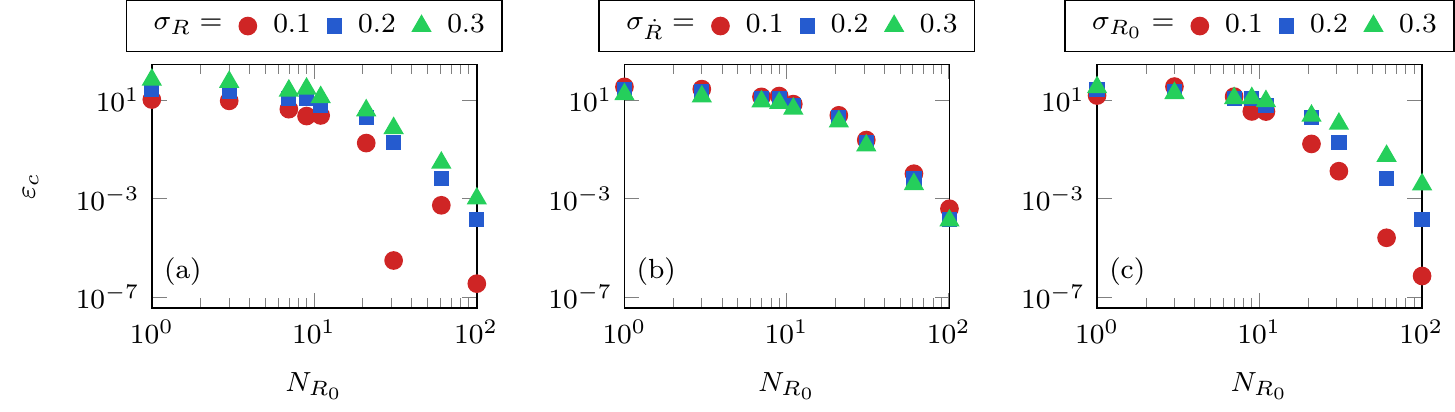}
    \caption{
        Relative closure error $\eps_c$ (defined in~\eqref{e:errorc}) for increasing number of $R_o$-direction quadrature points $N_{R_o}$.
        Variations in (a) $\sigma_R$, (b) $\sigma_{\Rdot}$, and (c) $\sigma_{R_o}$ are shown.
        Unless labeled otherwise, cases have the baseline $\sigma_R = \sigma_{\Rdot} = \sigma_{R_o} = 0.2$.
    }
    \label{f:Cerrors}
\end{figure}

Figure~\ref{f:Cerrors} shows the $N_{R_o}$ closure errors associated with variations in all three PDF directions (panels a--c).
For varying $\sigma_R$ (a) and $\sigma_{R_o}$ (c) we see that increasing variance $\sigma_{\cdot}$ results in larger closure errors $\eps_c$.
This appears to be associated with the larger pressure oscillations $p(x=0,t)$ observed for such cases.
For figure~\ref{f:Cerrors}~(b), we see the reverse trend, with larger $\sigma_{\Rdot}$ corresponding to smaller closure errors, though this effect is small.
Indeed, this effect matches that of the $R$- and $R_o$-direction effects, where larger $\sigma_{\Rdot}$ results in smoother pressure histories.
We also investigated the effect of $\Rey$, which results in smoother pressure profiles for smaller $\Rey$, and found the same trend.

\begin{figure}
    \centering
    \includegraphics{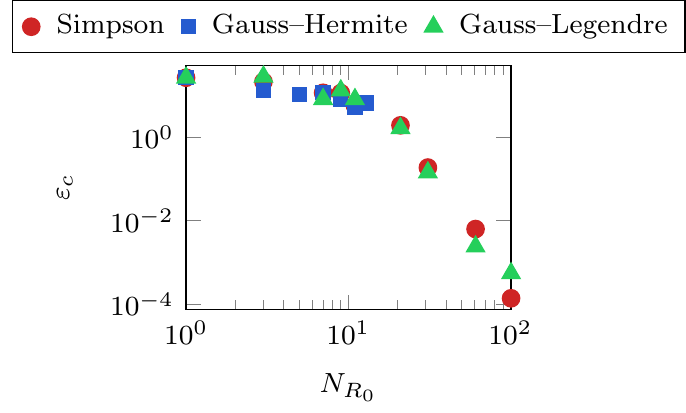}
    \caption{Closure error associated with $\sigma_R = \sigma_{\Rdot} = \sigma_{R_o} = 0.2$ for different $R_o$-direction quadrature rules as labeled.}
    \label{f:quadrules}
\end{figure}

Compelled by the closure error associated with $N_{R_o}$ dominating the total simulation error, we also implemented Gaussian--Hermite and Gauss--Legendre quadrature methods in the $R_o$-direction in an attempt to reduce the error.
Figure~\ref{f:quadrules} shows these closures errors $\eps_c$ for increasing $N_{R_o}$.
We see that these alternative quadrature rules do not significantly change the closure error, despite being optimal for a given $N_{R_o}$.
This is because the integrand becomes highly oscillatory without viscous damping, as bubbles of different equilibrium radii $R_o$ become out of phase~\citep{colonius08}.
Long-time simulations without significant viscous effects, broadly polydisperse bubble populations (large $\sigma_{R_o}$) will require treatment of these developing oscillations.
For example, a time-stationary analysis could be used to approximate the $R_o$-moments, so long as the bubble dynamics are fast compared to the fluid flow~\citep{colonius08}.
We will investigate this in future work.

\section{Conclusion}\label{s:conclusion}

A fully coupled numerical method for simulating sub-grid cavitating bubble dispersions was presented.
The approach represents and evolves the statistics of the bubble dynamic variables.
This work built upon the well-established framework of quadrature-based moment methods: a governing set of moment transport equations were derived, effectively representing the underlying statistics.
Those moments were inverted for a quadrature rule to close the fully coupled disperse flow equations.

Our results showed that only four quadrature points, two in each of the bubble dynamics coordinates, can be sufficient to represent the statistics of the monodisperse bubbles.
This model was particularly accurate for weak pressure forcing, and thus nearly linear bubble dynamics.
It was also accurate for highly damped dynamics, characteristic of low Reynolds numbers, as non-Gaussian statistics cannot present themselves.
This result contrasts against a previous approach that assumed Gaussian statistics, which had demonstrably worse performance and higher computational costs~\citep{Bryngelson2020}.
The method would require more quadrature points for cases with more complicated dynamics, such as larger forcing.

An acoustically excited dilute bubble screen problem demonstrated the model's behavior in a coupled flow.
Modeling the statistics in the bubble dynamics variables resulted in qualitatively different behavior in the screen region.
For example, increasing the distribution breadth in the bubble radius coordinate resulted in short-wavelength pressure oscillations of increasing magnitude superimposing the background response to the impinging acoustic wave.
Thus, modeling the $R$--$\Rdot$ distributions is potentially critical to modeling cavitating bubble clouds and certainly important if bubbles are ever in such statistical disequilibrium.

Finally, we found that for broadly polydisperse populations (large $\sigma_{R_o}$), the closure error in the equilibrium radius coordinate $R_o$ was most important.
Our results also show that phase-cancellation can modestly reduce some computational costs associated with resolving the $R_o$ coordinate.
Still, $R_o$-direction quadrature dominates the solution cost of these polydisperse bubble populations, even in the face of more sophisticated quadrature rules like Gauss--Hermite (corresponding to the underlying log-normal distribution).
However, one can apply existing approaches if time-separation is achieved, or a novel method based on, e.g., neural networks, can also treat this.
We will investigate these in-depth in future work.

\section*{Acknowledgements}

We thank Alexis Charalampopoulos and Themis Sapsis for invaluable discussions regarding the presented method.
The US Office of Naval Research supported this work under grant numbers N0014-17-1-2676 and N0014-18-1-2625.
Computations were performed via the Extreme Science and Engineering Discovery Environment (XSEDE) under allocations TG-CTS120005 (PI Colonius) and TG-PHY210084 (PI Bryngelson), supported by National Science Foundation grant number ACI-1548562.

\bibliographystyle{model1-num-names}
\bibliography{main}

\newpage
\begin{appendices}

\section{CHyQMOM moment inversion}\label{a:chyqmom}

This appendix describes the 2D, $2 \times 2$ (4) node CHyQMOM algorithm of~\citep{Fox2018}. 
We note that~\citet{patel19} provides the 3D version of this algorithm, though it is not necessary for our bubble dynamics problem.
Algorithm~\ref{al:chyqmom4} is the full 2D CHyQMOM algorithm, which references the 1D 2-node HyQMOM algorithm~\ref{al:hyqmom2}.
The optimal moment set for this case is
\begin{gather}
    \vbmom = \left\{ \mom_{00}, \mom_{10}, \mom_{01}, \mom_{20}, \mom_{02}, \mom_{11} \right\},
\end{gather}
which is also the input to algorithm~\ref{al:chyqmom4}.

\begin{algorithm}[H]
    \setstretch{\algostretch}
    \begin{algorithmic}[1]
        \Procedure{CHyQMOM4}{$\bmom$}
        \State $D_{ij} = \mom_{ij}/\mom_{00}$
        \State $C_{20} \leftarrow D_{20} - D_{10}^2$
        \State $C_{02} \leftarrow D_{02} - D_{01}^2$
        \State $C_{11} \leftarrow D_{11} - D_{10} D_{01}$
        \State $\brho,\bx^\prime = \textsc{HyQMOM2}(\{1,0,C_{20}\})$
        \State $\by^\prime = \bx^\prime C_{11} / C_{20}$
        \State $\mom_\mathrm{avg}^2 = C_{02} - \sum_j \rho_j {y_{j}^\prime}^2$ 
        \State $\tbrho, \tbx^\prime = \textsc{HyQMOM2}(\{1,0,\mom_\mathrm{avg}^2\})$
        \For{$i,j \in [1,2]$}
        \State $w_{ij} \leftarrow \mom_{00} \rho_i \trho_{j}$ 
        \State $x_{ij} \leftarrow D_{10} + x_j^\prime$
        \State $y_{ij} \leftarrow D_{01} + y_j^\prime + \tx_{i}^\prime$
        \EndFor \\
        \Return $\bw,\bx,\by$
        \EndProcedure
        \caption{
            CHyQMOM $2 \times 2$. 
            $\bmom$ are the input moments, $D$ are normalized moments, $C$ are central moments, and $\bw$, $\bx$ and $\by$ are the weights and node locations in the first and second coordinate directions (corresponding to $R$ and $\Rdot$ in the main text).
        }
    \label{al:chyqmom4}
    \end{algorithmic}
\end{algorithm}

\begin{algorithm}[H]
    \setstretch{\algostretch}
    \begin{algorithmic}[1]
        \Procedure{HyQMOM2}{$\bmom$}
        \State $C_2 \leftarrow (\mom_0 \mom_2 - \mom_1^2)/\mom_0^2$
        \State $w_1 = w_2 \leftarrow \mom_0 /2$
        \State $x_1 \leftarrow \mom_1 / \mom_0 + \sqrt{C_2}$
        \State $x_2 \leftarrow \mom_1 / \mom_0 - \sqrt{C_2}$ \\
        \Return $\bw,\bx$
        \EndProcedure
        \caption{
            The two-node 1D HyQMOM algorithm. 
            Nomenclature follows algorithm~\ref{al:chyqmom4}, with $\bw$ and $\bx$ serving as dummy weights and node locations.
        }
        \label{al:hyqmom2}
    \end{algorithmic}
\end{algorithm}

\section{CQMOM moment inversion}\label{a:cqmom}

This appendix has the same form as appendix~\ref{a:chyqmom}, though it instead describes the 2D CQMOM algorithm of~\citep{yuan11}. 
Algorithm~\ref{a:cqmom} is the full 2D CQMOM algorithm, referencing Wheeler's method for computing optimal 1D quadrature nodes and weights~\ref{al:wheeler}.
The optimal moment set for this case is
\begin{gather}
    \vbmom = \left\{ \mom_{00}, \mom_{10}, \mom_{01}, \mom_{20}, \mom_{02}, \mom_{11}, \mom_{30}, \mom_{03}, \mom_{12}, \mom_{13} \right\},
\end{gather}
which is also the input to algorithm~\ref{al:cqmom}.

\begin{algorithm}[H]
    \setstretch{\algostretch}
    \begin{algorithmic}[1]
        \Procedure{CQMOM}{$\bmom$}
        \State $\bw^{(x)}, \bx \leftarrow \textsc{Wheeler}(\{\mom_{00},\mom_{10},\dots,\mom_{2N_{x}-1,0}\})$
        \For{$i,j \in [1,\dots,N_x]$}
            \State $V_{ij} = x_j^{i-1}$
            \State $W_{ii} = w_i^{(x)}$
        \EndFor
        \For{$i \in [0,\dots,2N_{y}-1]$}
            \State $\bx_i^\prime \leftarrow \bV \bW \, \bx_i^\prime = \mom_{0:N_x-1,i}$
            \State $w_{ij}^{(y)}, y_{ij} \leftarrow \textsc{Wheeler}(\bx_i^\prime)$
        \EndFor
        \State $w_{ij} = w_i^{(x)} w_{ij}^{(y)}$ \\
        \Return $\bw,\bx,\by$
        \EndProcedure
        \caption{
            2D CQMOM algorithm with dummy coordinate direction $y$ conditioned on $x$.
            $\bV$ is a Vandermonde matrix and $\bW$ is a diagonal weight matrix.
        }
        \label{al:cqmom}
    \end{algorithmic}
\end{algorithm}

\begin{algorithm}[H]
    \setstretch{\algostretch}
    \begin{algorithmic}[1]
        \Procedure{Wheeler}{$\bmom$}
        \For{$i \in [1,\dots,2N]$}
            \State $\sigma_{1,i} \leftarrow \mom_{i-1}$
        \EndFor
        \State $a_0 \leftarrow \mom_1/\mom_0$
        \State $b_0 \leftarrow 0$
        \For{$i \in [2,\dots,N]$}
            \For{$j \in [i,\dots,2N-i+1]$}
            \State $\sigma_{i,j} \leftarrow \sigma_{i-1,j+1} - a_{i-2} \sigma_{i-1,i-1}$
            \EndFor
            \State $a_{i-1} \leftarrow \sigma_{i,i+1} / \sigma_{i,i} - \sigma_{i-1,i}/\sigma_{i-1,i-1}$
            \State $b_{i-1} \leftarrow \sigma_{i,i}/\sigma_{i-1,i-1}$
        \EndFor
        \For{$i \in [2,\dots,N-1]$}
        \State $J_{i,i} \leftarrow a_{i-1}$
        \State $J_{i+1,i} = J_{i,i+1} \leftarrow \sqrt{b_i}$
        \EndFor
        \State $\bQ \bLambda \bQ^{-1} \leftarrow \textsc{EVD}(\bJ)$
        \For{$i \in [1,\dots,N]$}
            \State $x_i \leftarrow \Lambda_{ii}$
            \State $w_i \leftarrow \mom_0 Q_{i,1}^2$ 
        \EndFor \\
        \Return $\bw,\bx$
        \EndProcedure
        \caption{
            Wheeler's algorithm for optimal quadrature points $\bx$ and weights $\bw$ given a moment set $\bmom$~\citep{Wheeler1974,marchisio13}.
            $\bJ$ is a Jacobi matrix, and EVD refers to the eigenvalue decomposition.
        }
        \label{al:wheeler}
    \end{algorithmic}
\end{algorithm}

\end{appendices}

\end{document}